\documentclass[aps,prx,twocolumn,superscriptaddress,nofootinbib,showkeys]{revtex4-1}

\usepackage{graphicx}
\usepackage{dcolumn}
\usepackage{bm}
\usepackage[utf8]{inputenc}
\usepackage{textcomp}
\usepackage{amsmath}
\usepackage[mathlines]{lineno}
\usepackage{siunitx}

\begin{document}

\title{Beam steering with a nonlinear optical phased array antenna}

\author{Sebastian Busschaert}
\author{Nikolaus Flöry}
\author{Sotirios Papadopoulos}
\author{Markus Parzefall}
\author{Sebastian Heeg}
\author{Lukas Novotny}

\affiliation{Photonics Laboratory, ETH Zürich, Zürich, Switzerland 
}

\date{\today}

\begin{abstract}
Transition metal dichalcogenides (TMDCs) exhibit high second harmonic (SH) generation in the visible due to their non-centrosymmetric crystal structure in odd-layered form and direct bandgap transition when thinned down to a monolayer. In order to emit the SH radiation into a desired direction one requires a means to control the phase of the in-plane nonlinear polarization.
Here, we couple the SH response of a monolayer MoS\textsubscript{2} to an optical phased array antenna and demonstrate controllable steering of the nonlinear emission. By exploiting the intrinsic SH generation by the phased array antenna we achieve uniform emission efficiency into a broad angular range. Our work has relevance for novel optoelectronic applications,  such as programmable optical interconnects and on-chip LIDAR. 
\end{abstract}

\maketitle

Nonlinear optical effects play a vital role in integrated photonics as frequency mixing and electro-optic modulation are necessary components for optical multiplexing and optoelectronic logic gates~\cite{Luo14,Reis12}. However, due to the generally weak nonlinear material responses, accessing these effects requires high incident powers. This has lead to an ongoing quest for materials with high intrinsic nonlinearities that can be incorporated into nano-scale devices. In addition, only non-centrosymmetric materials are capable of producing second order nonlinearities in the electric dipole approximation. Unfortunately, common materials such as silicon or gold are centrosymmetric and non-centrosymmetric materials such as lithium niobate are difficult to process on nanometer length-scales~\cite{Boyd08,Sutherland03,Siew18}.\\[-1.5ex]

Transition metal dichalcogenides (TMDCs) are a new class of materials that possess a strong nonlinear response and that can be structured with top-down nanofabrication techniques. These materials can be thinned down to monolayer form via mechanical exfoliation, which leads to the formation of a direct bandgap semiconductor~\cite{Mak10}. Any odd number of layers of TMDCs forms a non-centrosymmetric crystal and has been shown to exhibit high second-order nonlinearities, with reported values up to $\chi^{(2)} \sim 10^{-7}$m/V for the monolayer form~\cite{Kumar13,Malard13}. Second harmonic (SH) generation has been studied in detail for TMDCs and an extraordinarily high response was observed when the excitation or emission was chosen to be close to the optical bandgap of the respective material~\cite{Zhao16}.\\[-1.5ex]

Recently, several reports have studied the interaction between a TMDC monolayer and plasmonic particles and demonstrated an enhancement of the SH response~\cite{Wang18,Shi18}. The SH radiation can also be controlled by plasmonic metasurfaces~\cite{Chen18,Hu19} that shape the phase profile of the incoming light. For on-chip optoelectronic applications  it is desirable to steer the emission into desired directions and to have programmable control over the emission direction.\\[-1.5ex]

\begin{figure}[!b] 
\includegraphics{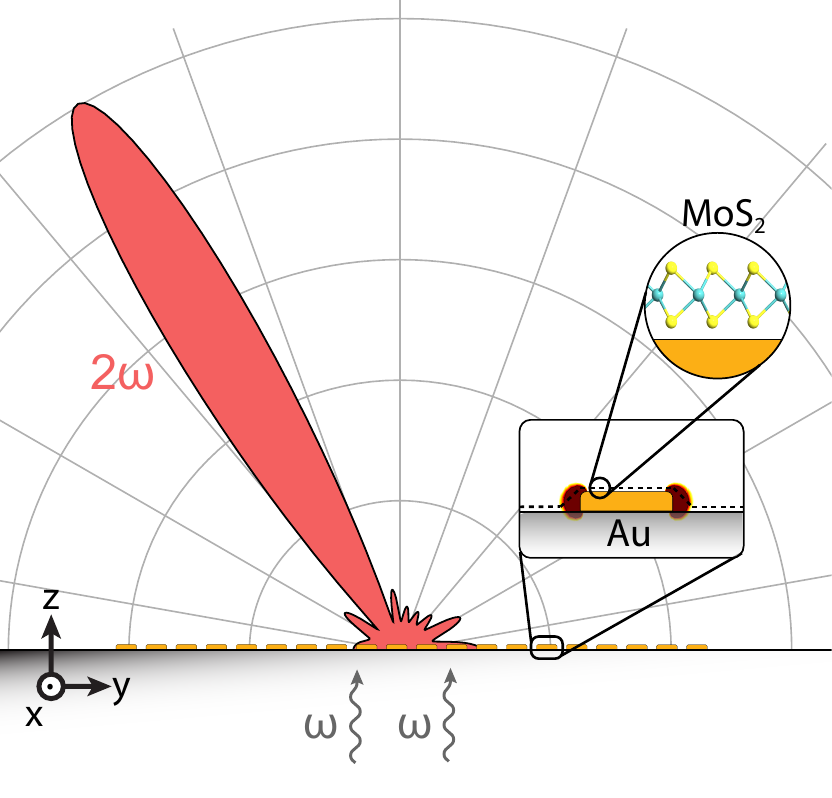}
\caption{Illustration of a MoS\textsubscript{2}-gold phased array antenna.}
\label{fig:wideTOC}
\end{figure}
As illustrated in Figure \ref{fig:wideTOC}, we demonstrate a MoS\textsubscript{2}-gold phased array antenna and demonstrate beam steering of SH radiation. We utilize the symmetry properties of the array and exploit the interplay between SH generated by MoS\textsubscript{2} and the phased array itself to generate a beam steering efficiency that is uniform over a broad angular range.\\[-1.5ex]

Our nonlinear phased array antennas were produced by the following procedure: First, arrays of equally-spaced gold nanorods were fabricated using electron-beam lithography. These arrays were then characterized by scanning electron microscopy (SEM) and by nonlinear spectroscopy. A monolayer of MoS\textsubscript{2} was then deposited on top of an array by using a dry transfer technique~\cite{Castellanos14,Jain18}. The deposited MoS\textsubscript{2} interacts with the local  optical near fields of the antenna elements and gives rise to SH generation. Nonlinear measurements before and after transfer of the MoS\textsubscript{2} on the same array allows us to quantify the nonlinear signal contributions.\\[-1.5ex]

\begin{figure}[!b] 
\includegraphics{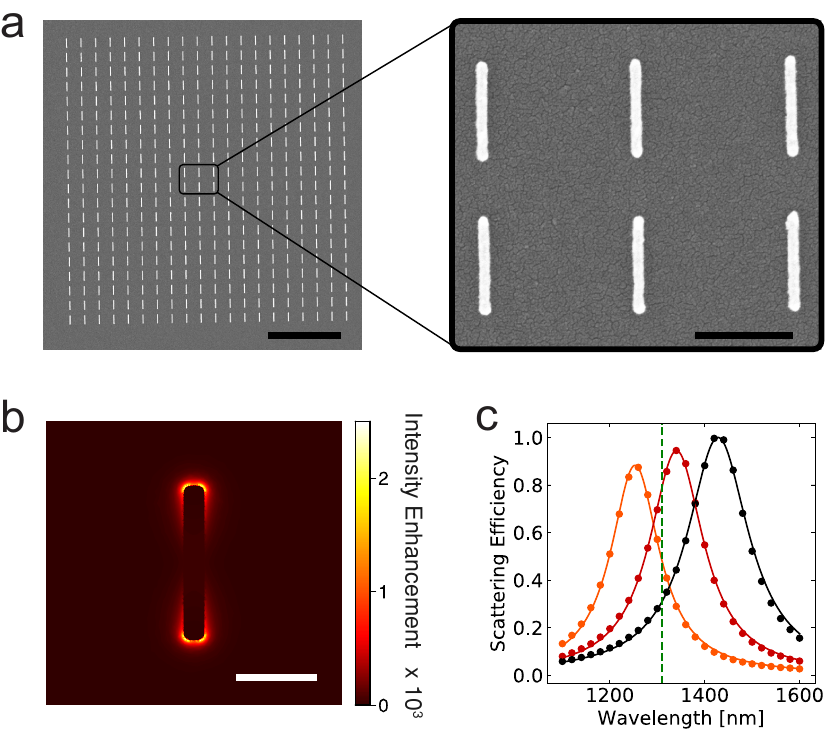}
\caption{(a) SEM image of a fabricated gold nanorod array on glass (scale bar:~\SI{2}{\um}) including zoomed-in image of the same array (scale bar:~\SI{250}{\nm}). The designed length was \SI{240}{\nm}. (b) Simulated intensity enhancement of a single gold nanorod (\SI[product-units = single]{220x30x20}{\nm\cubed}) for \SI{1310}{\nm} excitation, height of cut is \SI{10}{\nm}, scale bar: \SI{100}{\nm}. (c) Simulated length dependence of the localized surface plasmon polariton (LSPP) resonance of \SI{200}{\nm} (orange), \SI{220}{\nm} (red), \SI{240}{\nm} (black) long nanorods. The lines indicate Lorentzian fits and the green dashed line indicates the target wavelength. The width was fixed to \SI{30}{\nm} and the height to \SI{20}{\nm}.}
\label{fig:wide1}
\end{figure}
For our MoS\textsubscript{2}-array design we aimed towards maximizing the nonlinear response and obtaining the desired radiation pattern. In order to achieve a strong nonlinear response we made use of the plasmonic resonance of the gold nanorods constituting the phased array and the electronic resonance of MoS\textsubscript{2}. The particular dimensions of the rods were chosen such that the plasmonic resonance coincides with the target excitation wavelength at \SI{1310}{\nm}. This resulted in a SH wavelength of \SI{655}{\nm}, which, according to previous studies of MoS\textsubscript{2}, leads to a strong signal enhancement due to an electronic resonance associated with the A exciton~\cite{Zhao16}.\\ 
In Figure \ref{fig:wide1}a we show a scanning electron microscope (SEM) image of a fabricated gold nanorod array on glass and a magnified view. We employed three-dimensional finite element method simulations to calculate the scattering efficiency of the rods, as shown in Figures \ref{fig:wide1}b, c. In Figure \ref{fig:wide1}b we depict a plane-cut of the intensity enhancement of a simulated rod for a \SI{1310}{\nm} excitation wavelength. The strongest near field regions are at the ends of the nanorod, marking the desired interaction points with the MoS\textsubscript{2}. By simulating the scattering efficiency for rods with different lengths, shown in Figure \ref{fig:wide1}c, we demonstrate the length dependence of the plasmon resonance in the vicinity of our target wavelength (indicated by the green dashed line). Based on these results we fabricated samples with arrays of varying antenna lengths. In excellent agreement with the simulation we obtain the strongest nonlinear response from nanorod arrays of \SI{220}{\nm} length (see Supplementary Information), on which we will focus in the following. \\[-1.5ex]

The local field of each antenna element (rod) excites a SH response in the overlaid MoS\textsubscript{2} and in the rod itself. The SH radiation pattern of the antenna array follows from the coherent superposition of all SH signals.  For a phased-array of finite size, the array-characteristic emission pattern is described by the array factor~\cite{Milligan05,Balanis12}:
\begin{multline}\label{eq:1}
    \text{AF}(\phi_x,\phi_y) = \Bigg|\frac{1}{M^2}\frac{\sin{0.5M[  kd\sin{\phi_x}+\delta_x]}}{\sin{0.5[kd\sin{\phi_x}+\delta_x]}}  \\
    \times \frac{\sin{0.5M[kd\sin{\phi_y}+\delta_y]}}{\sin{0.5[kd\sin{\phi_y}+\delta_y]}}\Bigg|^2 ,
\end{multline}\\
where $M$ is the number of antennas in a column/row, $d$ is the distance between antennas, $k = 2\pi n_{\text{glass}} / \lambda_{\text{SH}}$ ($n_{\text{glass}}=1.52$) is the wave-vector at the SH wavelength in glass, $\phi_x$, $\phi_y$ are the emission angles and $\delta_x$,$\delta_y$ the phase gradients between antennas along $x$ and $y$, respectively. Generally, a larger number of antenna elements (large $M$) generates a sharper radiation pattern and the direction of the main radiation lobe can be set by the phase gradients $\delta_x$ and $\delta_y$. The existence (and number) of radiative grating orders is determined by the spacing $d$. We designed the antenna distance such that no grating orders exist for zero phase delay ($\delta_x=\delta_y=0$) but they can emerge for larger phase delays. This property is a distinct feature of the phased array and will be discussed later on. We investigated phased arrays with different array sizes $M$ and different distances $d$.  Here we discuss our results for an array with 20 x 20 elements and  distance of $d=400\,$nm.\\[-1.5ex]

\begin{figure*}[!t] 
\centering
\includegraphics{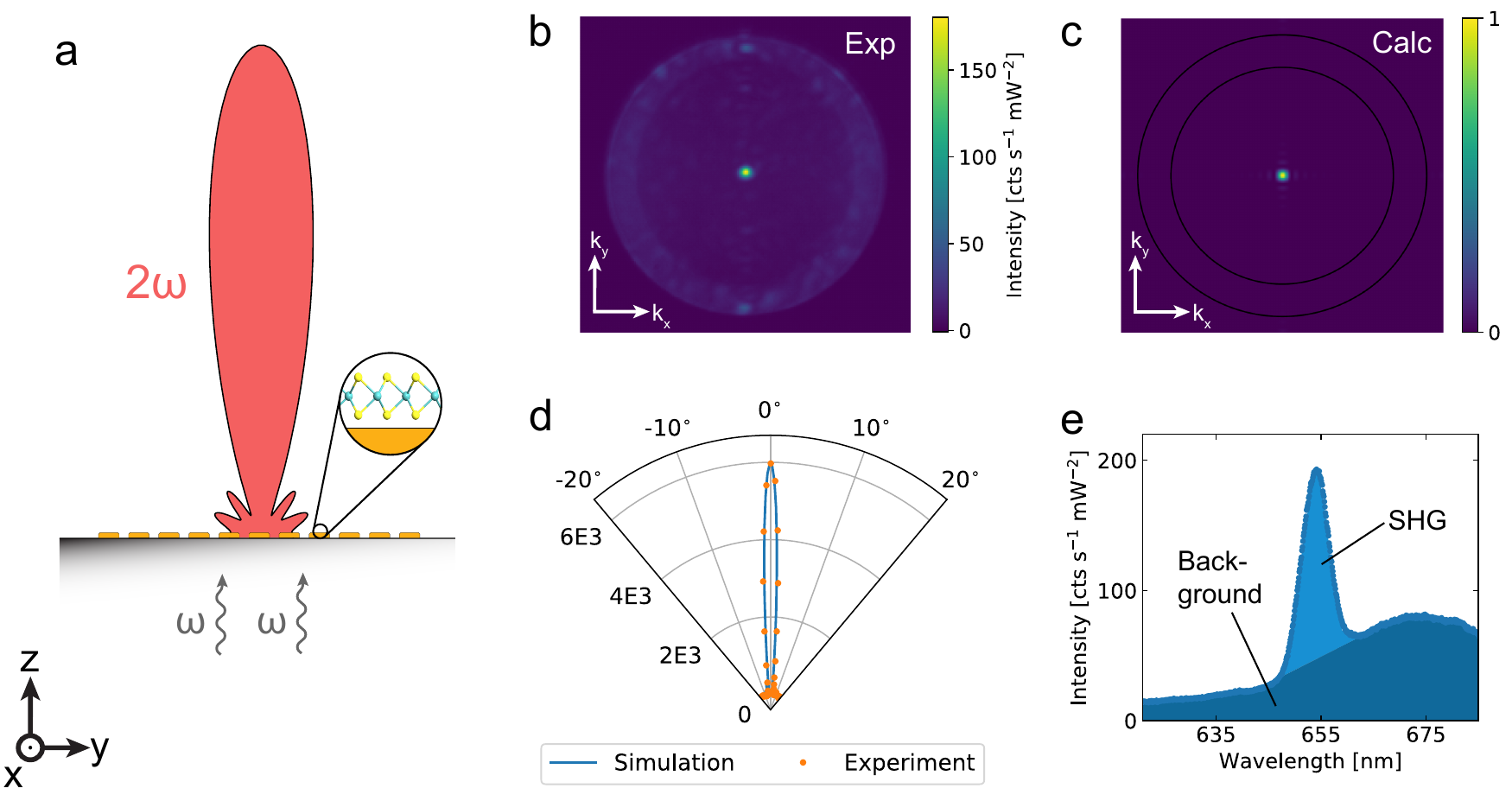}
\caption{Nonlinear measurements of the MoS\textsubscript{2}-array system under normal incidence. (a) Illustration of the studied system. \\  (b) Measured SH emission pattern of the coupled MoS\textsubscript{2}-array system. (c) Calculated SH emission pattern for 2D dipole array with $M = 20$. The dipoles are oriented along the $y$ direction. The black circles represent $k_\parallel = \sqrt{k_x^2+k_y^2} = k_0$ and $k_\parallel = \mathrm{NA} \times k_0$, with $\mathrm{NA} = 1.3$ being the numerical aperture of the objective. (d) Polar plot of the calculated (line) and measured (points) SH pattern along the intensity maximum. The intensity is given in units of cts s$^{-1}$ mW$^{-2}$ \SI{}{\micro\steradian}$^{-1}$. (e) Spectrum of SH emission (excitation at \SI{1310}{\nm}).}
\label{fig:wide2}
\end{figure*}
Nonlinear measurements were performed on the MoS\textsubscript{2}-array system in a reflection geometry setup. The array was illuminated with \SI{200}{\fs} pulsed laser light at \SI{1310}{\nm} through a 1.3 NA oil immersion objective. By focusing the laser onto the center of the back focal plane of the objective we obtain a collimated beam, resulting in an excitation with zero phase delay ($\delta_x = \delta_y = 0$). Optical filters were used to separate the SH emission from the back-reflected excitation.\\[-1.5ex]
 
The measured SH radiation pattern of the MoS\textsubscript{2}-array system is depicted in Figure \ref{fig:wide2}b. We observe a distinct peak at $(k_x,k_y)=(0,0)$, corresponding to emission normal to the sample surface, on top of a weaker, non-directional emission background. The distinct emission spot originates from the constructive interference of the coherent SH emission from individual antenna elements. In Figure \ref{fig:wide2}c we use Eq.~(\ref{eq:1}) to calculate the SH emission pattern as a function of $k_\parallel$, the in-plane wave-vector component, for the case of all array elements radiating in phase ($\delta_x = \delta_y = 0$). The measured pattern in Figure \ref{fig:wide2}b is in excellent agreement with the calculation. By taking cross-sections along $k_x$ for $k_y = 0$ and transforming the spatial coordinate into an angular representation we plot in Figure \ref{fig:wide2}d both the calculated and measured emission patterns as a polar plot. There is very good agreement between theory and experiment. We also record the spectrum of the emission from the MoS\textsubscript{2}-array system in a spectrometer (see Figure \ref{fig:wide2}e), which shows the SH signal at \SI{655}{\nm} on top of a spectrally broad background. The background is partially due to the multiphoton luminescence from gold ~\cite{Wang13,Wang15} and MoS\textsubscript{2}~\cite{Li15,Zhou17}. Due to its incoherent nature, the background is spectrally broad and non-directional, responsible for the weak background surrounding the sharp SH emission peak. \\[-1.5ex]

\begin{figure*}[!t]
\centering
\includegraphics{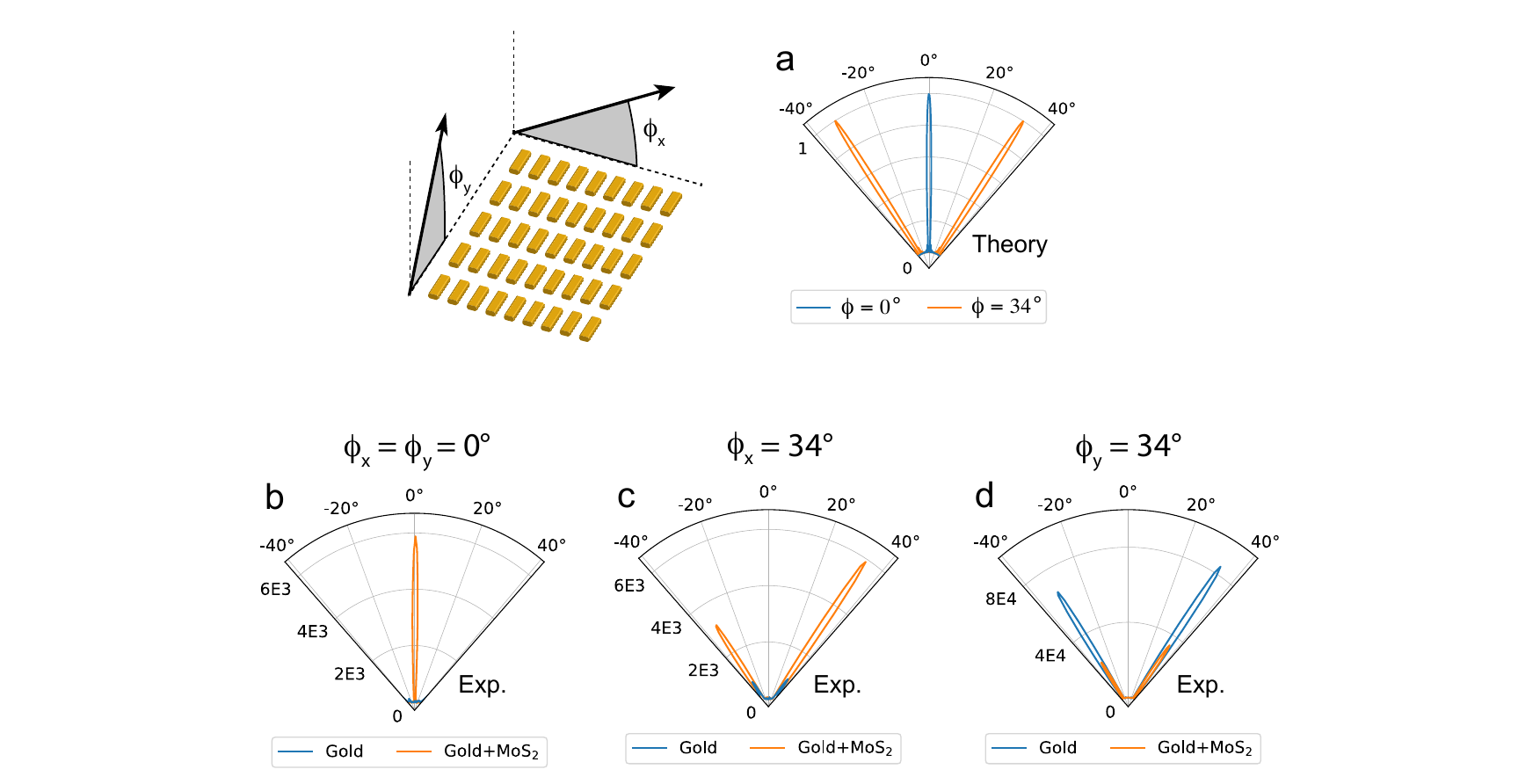}
\caption{Beam steering of SH emission by the phase delays along the $x$ direction ($\phi_x$) and the $y$ direction ($\phi_y$). The intensity of the measurements is given in units of cts s$^{-1}$ mW$^{-2}$ \SI{}{\micro\steradian}$^{-1}$. (a) Calculated emission patterns for zero phase delay (blue) and non-zero phase delay (orange) between antenna elements. (b) Measured SH emission of a bare gold array (blue) and a MoS\textsubscript{2}-array (orange) for zero phase delay. (c,d) Same as (b) but for a phase delay along $x$ yielding $\phi_x = 34^\circ$ (c) and for a phase delay along $y$ yielding $\phi_y = 34^\circ$ (d).}
\label{fig:wide4}
\end{figure*}
An important feature of a phased array antenna is the ability to steer the emission and control the change of emission characteristics, such as grating orders. Steering is achieved with the phase gradients $\delta_x$, $\delta_y$ in Eq.~(\ref{eq:1}), which define a phase delay between antenna elements. In Figure \ref{fig:wide4}a we illustrate the SH emission of a two dimensional dipole array for two different phase gradients, which results in emission into angles $\phi=0$ and $\phi=34^\circ$, respectively. Experimentally, we achieve a non-zero phase delay by laterally displacing the focus of the excitation beam in the back focal plane of the 1.3 NA oil immersion objective. This leads to a tilted wavefront and hence to a linear phase gradient. \\[-1.5ex]

In our experiments we use  MoS\textsubscript{2} for nonlinear conversion and the gold phased array to emit the SH radiation. It turns out, however, that there is a finite SH contribution from the gold phased array itself~\cite{McMahon06,Hooper19,Hamed19}. In order to distinguish between the two SH contributions we investigate the phased array antenna before and after transfer of MoS\textsubscript{2}.
Our experimental results show that for zero phase delay ($\delta_x=\delta_y=0$) the SH contribution from the gold array is negligible (Figure \ref{fig:wide4}b). The MoS\textsubscript{2}-array system exhibits the expected emission at $0^\circ$ whereas no signal is observed for the bare gold array. The same observation holds for phase delays along the $x$-axis (Figure \ref{fig:wide4}c), which corresponds to a tilt around the axis of a nanorod.
However, as shown in Figure \ref{fig:wide4}d, the SH contribution from gold becomes more prominent for large phase delays along the $y$-axis, that is, tilts perpendicular to the nanorod axis. We will analyze this situation later on. \\[-1.5ex]

Looking at the emission pattern shown in Figure \ref{fig:wide4}c, we note that there are two emission peaks, one at $34^{\circ}$ (main emission peak) and the other at $-34^{\circ}$ (first grating order). Both of them are predicted by Eq.~(\ref{eq:1}) and indicate the interaction between the gold nanorod array and the MoS\textsubscript{2}. In order to quantify the interaction strength we compare SH generation from the coupled MoS\textsubscript{2}-array system with SH generation from a bare MoS\textsubscript{2} monolayer (see Supplementary). We carry out this comparison for $\delta_x = \delta_y = 0$, as the SH contribution from the gold array can be neglected in this case, cf. Figure \ref{fig:wide4}b. Comparing the absolute intensities of the MoS\textsubscript{2}-array system and the bare MoS\textsubscript{2} only, we determine an enhancement of $E = I_{\text{MoS}_2\text{-array}}/I_{\text{MoS}_2} \approx 1.6$. This value denotes the area enhancement, that is, the enhancement averaged over many antenna elements. Note that the local signal enhancement at the poles of the nanorods is much larger (see Figure \ref{fig:wide1}b).\\[-1.5ex]

In Figure \ref{fig:wide4}d we apply a phase delay along the $y$ direction which coincides with the nanorod axes. In contrast to the previous case, we now observe a significant SH response from the bare gold array. We thus find that the SH signal emitted in the ($y$,$z$) plane is dominated by the gold nanorods whereas the SH signal in the ($x$,$z$) plane originates from MoS\textsubscript{2}. This is a favorable finding as it allows us to emit SH radiation in different directions with similar efficiency.\\[-1.5ex]

\begin{figure}[!t]
\includegraphics[width=0.41\textwidth]{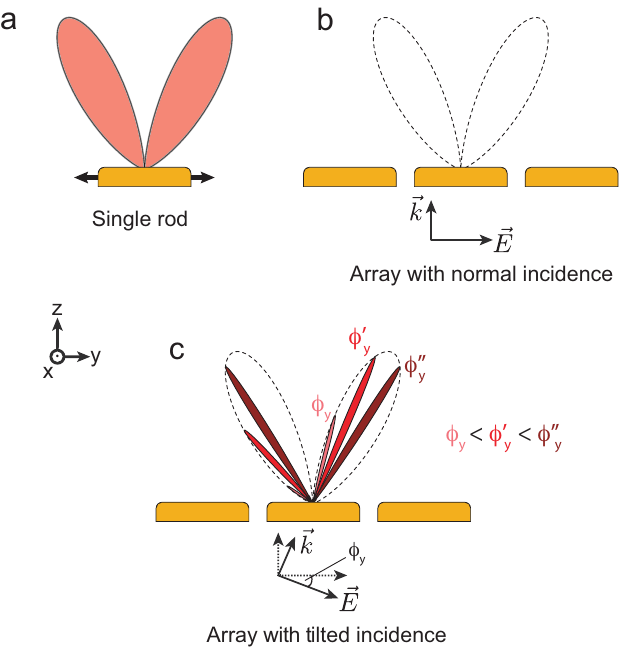}
\caption{Symmetry considerations for SH generation from a bare gold nanorod array. (a) Long axis view of single nanorod SH emission (only upper half space shown), indicated as red lobes. The emission is symmetric with respect to the z axis. The black arrows denote electric dipoles. (b) Long axis view of a nanorod array under normal incidence. Destructive interference cancels the SH emission. (c) Long axis view of a nanorod array under tilted incidence (tilt angle $\phi_y$).}
\label{fig:wide3}
\end{figure}
Let us analyze why the gold array generates significant SH radiation only along one tilt axis. Although gold exhibits no bulk SH generation due to its crystal structure, symmetry breaking at surfaces and edges can lead to a non-zero SH contribution~\cite{Bouhelier03,Bachelier10,Reichenbach12,Kauranen12}. The SH emission pattern of a single nanorod is generated by the electric near fields at the ends of the rod (cf. Figure \ref{fig:wide1}c). The resulting dipoles formed at opposite ends of the rod possess a phase difference of $\pi$, such that their radiation destructively interferes in the farfield at an emission angle of 0$^\circ$. The two out-of-phase dipoles give rise to a double-lobed SH emission pattern (Figure \ref{fig:wide4}a) which, however, cancels out by destructive interference when multiple nanorods are lined up in an array and radiate in phase (Figure \ref{fig:wide4}b). If the incoming beam is tilted such that the main emission lobe is no longer normal to the surface, the destructive interference is lifted (Figure \ref{fig:wide3}c). For this case SH generation is no longer dipole forbidden for observation angles in the (y,z) plane.\\[-1.5ex]

\begin{figure*}[!t] 
\centering
\includegraphics[width=0.93\textwidth]{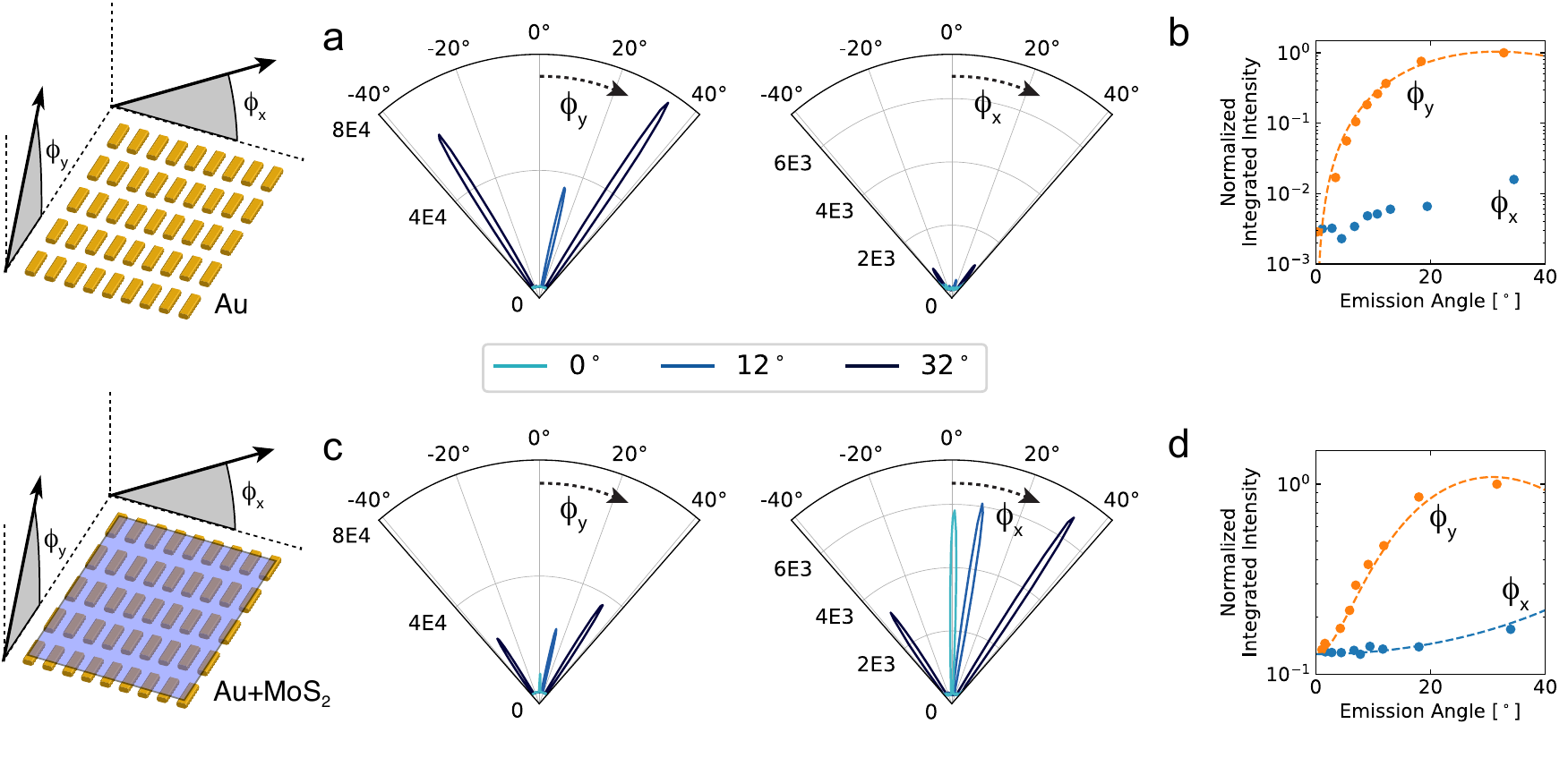}
\caption{Comparison of SH generation for different phase delays between a bare gold array and a MoS\textsubscript{2}-array system. (a) Measured SH emission patterns of a bare gold array for three different phase delays along the $y$-direction (left) and the $x$-direction (right). The intensity is given in units of cts s$^{-1}$ mW$^{-2}$ \SI{}{\micro\steradian}$^{-1}$. (b) Integrated SH intensities as a function of emission angle. The values are normalized with the value measured at $\phi_y=34^\circ$. The dashed orange line corresponds to a theoretical model, as explained in more detail in the main text. (c,d) Corresponding results for the combined MoS\textsubscript{2}-array system. The emission efficiency is more uniform for the case of the MoS2-array system.}
\label{fig:wide5}
\end{figure*}
In order to confirm the explanation described in Figure \ref{fig:wide3} we changed the phase delay along the two symmetry axes in small increments and tracked the SH intensity of the main emission lobe. In Figure \ref{fig:wide5}a we show the behavior of the bare gold array, first for the $y$-direction ($\phi_y$) and then the $x$-direction ($\phi_x$). For zero phase delay ($\delta_x=\delta_y = 0$) we do not observe any significant SH contribution, however as we increase the phase delay along $y$ the SH emission appears and increases with increasing phase delay. For the $x$ direction, on the other hand, there seems to be only a small, negligible SH contribution (note the different scales in the plots). The integrated values for several emission angles are compiled in Figure \ref{fig:wide5}b and normalized to the highest measured SH value. The values along $x$ are two orders of magnitude smaller than along $y$ and remain constant until the maximum applied phase delay. In order to understand the increase of SH generation along $y$ we consider two dipoles at opposite ends (as illustrated in Figure \ref{fig:wide3}a) and calculate the radiation pattern of a single nanorod as a function of solid angle $I_{\text{quad,em}}(\phi)$. We account for the fact that the local excitation intensity is angle-dependent (due to the array) $I_{\text{exc}}(\phi)$. The same is true for the transmission of the excitation through the sample $I_{\text{trans}}(\phi)$ (for more details see Supplementary). We then obtain the angle-dependent SH intensity of a single nanorod as\\[-1.5ex]

\begin{equation}\label{eq:2}
    I_{\text{antenna}}(\phi) = I_{\text{quad,em}}(\phi) I_{\text{exc}}(\phi) I_{\text{trans}}(\phi)  
\end{equation}\\
The result of this calculation is shown as a dashed line in Figure \ref{fig:wide5}b and agrees well with the data points.\\
In order to obtain the complete SH emission pattern of the full array, the expression has to be multiplied with the array factor in Eq. (1) \\
\begin{equation}\label{eq:3}
\centering
    I_{\text{phased array}}(\phi) = \text{AF}(\phi) I_{\text{antenna}}(\phi)  
\end{equation}
For the MoS\textsubscript{2}-array system we are not only dealing with the nonlinearity of the bare gold array (which we denote by $I_{\text{quad,em}}(\phi)$) but also by the nonlinearity of the MoS\textsubscript{2}. The same analysis as for the bare gold array was conducted for the MoS\textsubscript{2}-array system and the results are shown in Figures \ref{fig:wide5}c and d. Compared to the bare gold array, the overall signal strength is weakened for a phase delay along the $y$-direction ($\phi_y$) but the qualitative behavior is similar. For phase delays along the $x$ direction we observe that the SH signal is constant for a wide range of angles $\phi_x$ and only shows a small increase for a large $\phi_x$. This observation agrees with the explanation given in Figure \ref{fig:wide3}, which predicts that the emission along the $x$ direction is solely due to SH generation by MoS\textsubscript{2}.\\[-1.5ex]

In Figure \ref{fig:wide5}d we show the integrated SH signal as a function of emission angle. We observe that the signal strengths along the two symmetry axes differ only by one order of magnitude (for the bare gold array the difference was nearly three orders of magnitude, c.f. Figure \ref{fig:wide5}b). This favorable behavior arises from the fact that both MoS\textsubscript{2} and the gold array contribute to the nonlinear signal generation. In order to describe our measured data we use Eq.~(\ref{eq:2}) with the addition of a dipolar emission pattern that describes the response of the MoS\textsubscript{2} (see Supplementary):\\
\begin{multline}\label{eq:4}
    I_{\text{antenna}}(\phi) = |E_{\text{quad,em}}(\phi)+E_{\text{dipole,em}}(\phi)|^2 \\ \times I_{\text{exc}}(\phi)I_{\text{trans}}(\phi) 
\end{multline}\\
The resulting theoretical curve agrees well with the data points and explains the two different SH contributions (MoS\textsubscript{2} and gold array) in the measured emission patterns.\\ [-1.5ex]
 
\begin{figure}[!b]
\centering
\includegraphics{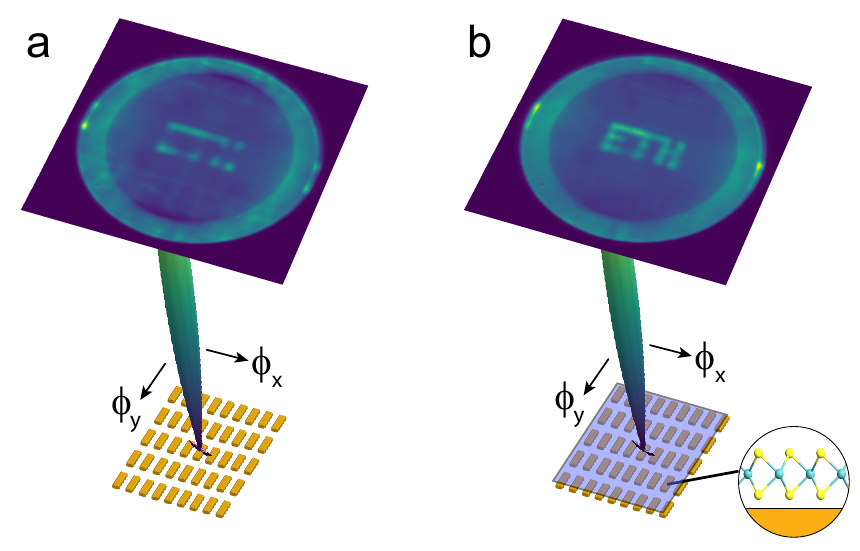}
\caption{Dynamic beam steering of SH emission. The phase delays along $\phi_x$ and $\phi_y$ were varied in time to write a "ETH" farfield pattern. (a) Resulting pattern for the bare gold array. (b) Resulting pattern for the MoS\textsubscript{2}-array system.}
\label{fig:wide6}
\end{figure} 
We finally use the MoS\textsubscript{2}-array system for dynamic beam steering and compare it with the bare gold array. We used a motorized lens-system that tilts the incoming wavefront in the sample plane, thereby dynamically changing the applied phase delay. Using this approach we projected the pattern "ETH" into the farfield. This pattern, shown in Figure \ref{fig:wide6}, was imaged with a CCD camera with an integration time of  50 seconds. As demonstrated in Figure \ref{fig:wide6}a, writing with the bare gold array results in all horizontal writing lines missing due to the symmetry forbidden region. However, when we write the same sequence with the MoS\textsubscript{2}-array system, shown in Figure \ref{fig:wide6}b, we profit from SH generation from MoS\textsubscript{2}. The resulting farfield pattern now shows the full "ETH" logo, including all vertical and horizontal lines. Thus, the combined MoS\textsubscript{2}-array system allows us to steer SH radiation into arbitrary directions. \\[-1.5ex]

In conclusion, we showed the nonlinear coupling of a monolayer semiconductor, here MoS\textsubscript{2}, to an array of gold nanorods. The combined system functions as a phased array antenna that emits SH radiation into arbitrary directions. Beam steering is accomplished by controlling the phase delay between antenna elements along the two main axes. The emission efficiency is not only dictated by the array factor, but also by the emission pattern of a single antenna element. The presented results offer a new platform to study and understand the coupling of second order nonlinear effects mediated by TMDCs and structured surfaces. Furthermore, nonlinear optical phased array antennas can have applications for dynamical optical interconnects and on-chip LIDAR.\\

\section*{Methods}

\subsection*{Electrodynamic Simulations}
Numerical simulations were carried out with the wave optics module of COMSOL Multiphysics. A single gold rod with dimensions of 30 nm width, 20 nm height and variable length was defined on top of a glass substrate (n = 1.52), whereas the rod was surrounded by air (n = 1) in the other half space. For the dielectric function of gold we used measured values~\cite{Johnson72}. The simulations were carried out in two steps: First, the fields of the geometry were simulated under plane wave illumination without rods such that the background field could be determined. In a a second step, the geometry was simulated with rods. The scattering efficiency of a single rod was calculated by normalization with the cross-section of the rod. The two half spaces were surrounded by perfectly matched layers (PMLs) to exclude back-reflection. Comparing the resonance condition of a single rod with the measured resonance of the full array did not show any significant deviations, which is due to the large separation between antenna elements (negligible near field interaction between antenna elements).

\subsection*{Sample Fabrication}
Phased array antennas were fabricated on commercial glass coverslips. First, markers were patterned by means of electron beam lithography (EBL), evaporation of 5\,nm titanium (Ti, 0.05\,nm/s) and 50\,nm gold (Au, 0.2\,nm/s) at a pressure of $<$\,10$^{-7}$\,mbar and subsequent lift-off (acetone, ispropyl alcohol and deionized water). The antenna arrays consisting of Au nanorods were then defined again by EBL as single pixel lines (SPL, line-dose 1400\,pAs/cm), followed by development and soft plasma cleaning (O$_2$ plasma, 70\,W, 30\,s). 20\,nm Au was deposited by electron beam evaporation (0.1\,nm/s) at a pressure of $<$\,10$^{-7}$\,mbar, followed by lift-off. For scanning electron microscopy a sample fabricated with the same parameters was sputter-coated with \SI{1.5}{\nm} Pt/Pd and then imaged using the in-lens detector (TLD) of a FEI Magellan 400 system.\\[-1.5ex]

For the final structure including the monolayer MoS\textsubscript{2}, bulk MoS\textsubscript{2} crystal was exfoliated onto UV-ozone cleaned polydimethylsiloxane (PDMS). After observing a monolayer via optical contrast difference, the respective flake was stamped on top of a suitable array with \SI{}{\um} precision using a SUSS MicroTec mask aligner.

\subsection*{Nonlinear Measurements}
For the nonlinear measurements we used the IR output of an optical parametric oscillator (Coherent Mira-OPO) that provided 200 fs pulses at a repetition rate of 76 MHz. All experiments were conducted with 1310 nm light, such that both SH generation and TH generation were detectable in the visible light spectrum. To exclude any pump contributions we spectrally filtered the laser (Semrock BL 1110LP, Chroma HHQ940LP). The beam was sent into a 90\textdegree-periscope consisting of two motorized stages (PT1/M-Z8 by Thorlabs) and an attached lens (AC254-400-C-ML by Thorlabs), that enabled us to focus the laser onto the back focal plane (BFP) of an oil immersion objective (1.3 NA Plan Fluor Nikon) and to laterally displace the beam. Focusing the laser onto the BFP leads to a large area illumination of the sample and lateral displacement of the beam gives rise to a wavefront tilt and hence a phase delay between antenna elements. With an initial beam waist of 2 mm, the beam spot diameter in the sample plane was roughly \SI{10}{\um}. The average power before entering the periscope was determined to be 1.6 mW. The excitation peak intensity never exceeded 1 GW/cm$^2$, which is below the damage threshold for the gold nanorods. The incident polarization was chosen to be along the long axis of the nanorods. \\[-1.5ex]

The excitation and emission wavelengths were seperated by a dichroic beamsplitter (DMSP950R by Thorlabs), further spectral selection was performed via optical filters (770SP, 650/60BP for SH generation, 561SP for THG, all filters by Semrock) in front of the detectors. The emission was sent either to a spectrometer (Acton SP2300 with Pixis100 CCD) or an EMCCD (Acton Photon Max 512). Different lens arrangements for real space and Fourier space imaging. \\[-1.5ex]

For calibration of the emission angles we used the first concentric ring in the BFP (NA=1 or $\phi=41.14^\circ$) for normalization. The error in determining the angle is $\pm 1^\circ$ (see Supplementary Information for more details).

\begin{acknowledgments}
This work was financially supported by the Swiss National Science Foundation (grant no. 200021\_165841). S. Heeg acknowledges financial support by ETH Zurich Career Seed Grant SEED-16 17-1. The authors thank A. Jain for valuable input during sample fabrication and D. Windey for providing support during data analysis. Furthermore, the authors acknowledge the use of the facilities of FIRST center of micro- and nanoscience and ScopeM at ETH Zürich. \\
\end{acknowledgments}

\nocite{*}
\providecommand{\noopsort}[1]{}\providecommand{\singleletter}[1]{#1}%

\end{document}